\documentclass[12pt,a4paper]{cibb}

\makeatletter
\providecommand{\@ordinalM}[2]{#1}
\makeatother

\usepackage{subfigure,graphicx}
\usepackage{amsmath,amsfonts,latexsym,amssymb,euscript,xr}
\usepackage{booktabs}
\usepackage[nodayofweek]{datetime}
\usepackage{hyperref}
\usepackage{fmtcount}
\usepackage[english]{datenumber}
\usepackage[absolute]{textpos}

\usepackage{xurl}
\usepackage[table]{xcolor}
\usepackage{color,colortbl,tabularx}

\usepackage[english]{babel}
\usepackage[protrusion=true,expansion=true]{microtype}
\usepackage{amsmath,amsfonts,amsthm}
\usepackage{pifont}

\def\black{\color{black}}
\def\blue{\color{blue}}

\newcommand{\review}[1]{\textcolor{black}{#1}}

\definecolor{LightBlue}{rgb}{0.88,0.9,0.9}

\title{\Large $\ $\\ \bf Leveraging hologenomic data for phenotypic prediction: potential and pitfalls}

\author{\blue \large Solène Pety$^{*,1,2}$, Ingrid David$^{3}$, Andrea Rau$^{1}$ and Mahendra Mariadassou$^{2}$}

\address{\blue \footnotesize $\ $\\$^1$ Université Paris-Saclay, INRAE, GABI, 78350, Jouy-en-Josas, France \\
$^2$ Université Paris-Saclay, INRAE, MaIAGE, 78350, Jouy-en-Josas, France \\
$^3$ Université de Toulouse, INRAE, ENVT, GenPhySE, 31326, Castanet-Tolosan, France  \\

\bigskip
ORCID codes: SP 0000-0002-1245-7425; ID 0000-0002-2514-6693; AR 0000-0001-6469-488X; MM 0000-0003-2986-354X
\bigskip
\newline
$^*$corresponding author: solene.pety@inrae.fr
\vspace{-2mm}
}

\abstract{\small holobiont, multi-omics integration, compositional data, multi-generation \normalsize
\\[17pt]
{\bf Abstract (max 250 words).} The microbiota is increasingly recognized as an active component of host biology, influencing various host phenotypes. Advances in high-throughput sequencing and the emergence of the holobiont perspective have raised expectations regarding hologenomic-informed prediction. Yet, whether and under which conditions integrating microbiota and genomic data meaningfully improves phenotypic prediction remains unclear. The biological characteristics of the microbiota, including but not limited to transmission mechanisms, environmental effects and interactions with host genetics, complicate their integration into classical evaluation frameworks. In addition, microbiota datasets are high-dimensional, highly dispersed, sparse and compositional.  Finally, analytical choices such as the taxonomic granularity considered for aggregation or the similarity matrix used in prediction models may impact downstream inference and prediction accuracy. Here we explore these challenges using a comprehensive set of transgenerational hologenomic simulations. By generating controlled and contrasted biological scenarios across a broad parameter space, we examine how microbiota granularity, variance structure and host modulation influence (i) the estimation of variance components and (ii) the accuracy of phenotypic prediction. We show that the added value of hologenomic, compared to genomic prediction, is highly context dependent. Our results provide a structured framework to interrogate when and how integrating microbiota may enhance phenotypic prediction in breeding applications.
}

\begin{document}

\renewcommand{\thefootnote}{}
\footnotetext{\small{Article version: \datedate $\;$ h\currenttime  $\;$ CET}}

\thispagestyle{myheadings}
\pagestyle{myheadings}
\markright{\tt Proceedings of CIBB 2026}

\section{Introduction}
\label{sec:SCIENTIFIC-BACKGROUND}

The microbiota refers to the community of microorganisms within a host, 
constituting a complex and dynamic ecosystem shaped by interactions between the environment and host-related factors. Through diverse metabolic pathways and host–microbe interactions, microorganisms influence a wide range of complex traits of interest (\textit{e.g.} nutrient assimilation, stress resilience, methane emission, \review{host immune response} or growth performance) ~\cite{forcinaGutMicrobiomeStudies2022}.
The rapid development of high-throughput sequencing technologies has profoundly transformed microbiome research over the last decade, 
enabling the large-scale characterization of microbial communities and leading to the holobiont perspective, in which host and microbiota are considered to be an integrated functional unit ~\cite{zilber-rosenbergRoleMicroorganismsEvolution2008a}. These advances have progressively positioned the microbiota as a potential complementary lever to genetics in breeding programs, fueling expectations regarding microbiota-informed prediction of phenotypes in both plant and livestock systems ~\cite{baiEffectsProbioticsupplementedDiets2013}. The biological features of the microbiota, such as transmission mode, environmental sensitivity and interactions with host genetics, substantially influence the observable signal. In addition, microbiota datasets are high-dimensional, highly dispersed, sparse and compositional. Thus, analytical choices, including taxonomic aggregation level and the approach used to characterize microbiota-based similarity among individuals, may interact with the underlying biological scenario, making it difficult to disentangle methodological artefacts from biological effects ~\cite{nearingMicrobiomeDifferentialAbundance2022}.  

To address this, we generated a large set of controlled biological scenarios in which parameters of interest such as microbiability $b^2$ (the fraction of phenotypic variance explained by the microbiota), direct heritability $h^2_d$ (the fraction of phenotypic variance explained by genetics not mediated by microbiota), transmission dynamics and genetic modulation are explicitly defined using RITHMS ~\cite{petySimulatingTransgenerationalHologenomes2025}, a flexible framework for transgenerational hologenomic simulations. This simulation-based approach enabled us to examine how analytical choices and biological contexts jointly shape variance component estimation and phenotypic prediction using hologenomic data.

\section{Data and Methods}
\label{sec:DATA-AND-METHODS}

We consider the context of a multi-generation livestock population in which both genomic information and microbiota composition are available for each individual. We focus on the case where a phenotype of interest has been recorded on all individuals with the exception of the selection candidates (\emph{i.e.} individuals in the final generation),
as is the case for traits measured in adults such as milk production or longevity. We then aim to evaluate whether hologenomic data can improve phenotypic prediction over microbiota or genomic data alone. In this case, partitioning phenotypic variance across genetic and microbiota components is useful for understanding the added predictive value of microbiota data.

\subsection{\textbf{Simulating transgenerational hologenomic data}}

We used a founding population of 750 pigs with both genomic data (here, using a subset of 5000 SNPs from a 70K SNP GeneSeek GGP Porcine HD Chip) and gut microbiota \review{(taxonomic profiles based on Illumina MiSeq sequencing of the V3-V4 region of the 16S rRNA gene from fecal samples and filtering out taxa with prevalence $\leq$ 5\% after rarefaction,  see~\cite{deruGeneticRelationshipsEfficiency2022} for details)}. We then used RITHMS to simulate genotypic, microbiota and phenotypic data over 5 generations. 

Genomic data simulation uses the standard approach described in ~\cite{pookMoBPSModularBreeding2020}. We focus here on the simulation details of the microbiota and phenotypic data. RITHMS models microbiota dynamics across generations through two biological processes: vertical transmission (from dam to offsprings) of microorganisms at rate $\lambda$ and horizontal transmission (from animals coliving in the environment) at rate $1 - \lambda$. Phenotypes were simulated at each generation as a linear combination of QTL and causal-taxon effects, where the abundances of the latter are submitted to host genetic control (\review{parametrized} by $\sigma_\beta$, where larger values correspond to stronger control). Phenotypic data in the 5th generation were masked from model fitting to evaluate prediction accuracy for selection candidates.



\subsection{\textbf{Key analytical steps for microbiota data in hologenomic prediction}}

Integrating microbiota data into a phenotypic prediction framework involves a sequence of analytical steps, each of which can substantially influence downstream results. We focus here on two aspects involving the microbiota that are directly impacted by these analytical decisions. Integrating microbiota data into a genomic prediction framework involves a sequence of analytical decisions, each of which can substantially influence the output accuracy. We structure this pipeline into three main steps and identify the two points most impacted by analytical steps involving the microbiota.

\subsubsection{Microbiota aggregation level}

Microbiota abundance data are high-dimensional count matrices, typically sparse, highly dispersed and compositional. Compositionality and sparsity require dedicated preprocessing, such as the centered log ratio (CLR) transformation, and special handling of zeroes, such as the addition of pseudo-counts. The resolution chosen for aggregating taxa is also a crucial choice. In practice, the finest resolution is not always the most relevant: amplicon sequence variants (ASV) or operational taxonomic units (OTU) can be very host-specific, whereas  phenotypic effects may be conserved to a point at higher taxonomic levels (genus, family, etc). As such, considering microbiota data at the finest resolution may lead to a lack of generalizability, while overly coarse resolutions may lose signal leading to a lack of power.

\subsubsection{Model fitting}

The dominant approach used for phenotypic prediction is the linear mixed model, in which the relationships between individuals are quantified using a similarity matrix built from pre-processed genotypes and/or abundances. In the latter case the taxonomic resolution directly conditions the information available to the model. In the genomic context, the GBLUP framework has been extended to omics data by replacing or augmenting the genomic relationship matrix with analogous omic-specific kernels ~\cite{legarraGenomicEvaluationMethods2023}, as shown in Equation~ \ref{eq:mgblup}.
\begin{equation}\label{eq:mgblup}
    \underbrace{\underbrace{\boldsymbol{y} = \boldsymbol{\mu} + \boldsymbol{g} + \boldsymbol{\epsilon}}_{\text{GBLUP}}
    + \boldsymbol{b}}_{\text{MGBLUP}}
\end{equation}
where $\boldsymbol{y}$ is a vector of phenotypes, $\boldsymbol{\mu}$ is the vector of shared intercepts, $\boldsymbol{g} \sim \mathcal{N}(0, \mathbf{K_G})$  the genetic effects, $\boldsymbol{b} \sim \mathcal{N}(0, \mathbf{K_B})$ the microbiota effects and $\boldsymbol{\epsilon}$ the normally distributed residuals. The kernels are given by  $\mathbf{K_G} = \boldsymbol{G}\boldsymbol{G}^T/ n_g$, where $\boldsymbol{G}$ is a matrix of centered and standardized genotypes and $n_g$ is the number of variants, and similarly $\mathbf{K_B} = \boldsymbol{B}\boldsymbol{B}^T/ n_b$, where $\boldsymbol{B}$ is the pre-processed CLR-transformed abundance matrix aggregated at the taxonomic resolution described in the previous step and $n_b$ is the number of (potentially aggregated) microorganisms. Joint models incorporating both genomic and microbiota kernels (MGBLUP) allow the variances attributable to each source to be estimated simultaneously.  In this work, we fitted GBLUP and MGBLUP models using the BGLR package. 
Beyond kernel-based approaches, other methods such as Bayesian sparse models or regularised regressions offer complementary ways to handle microbiota data. 

\subsection{\textbf{Evaluating contrasted biological scenarios}}

\review{We explored two complementary sets of scenarios. First, to investigate the effect of taxonomic granularity, we varied the aggregation level of microbiota data while fixing $h^2_d = b^2 = 0.25$. Second, to characterise the conditions under which microbiota integration improves phenotypic prediction, we jointly varied the vertical transmission rate $\lambda$, the microbiability $b^2$, and the strength of genetic control on microbiota composition $\sigma_\beta$, with a fixed direct heritability $h^2_d = 0.2$. These parameters were chosen because they directly govern the amount of independent information that microbiota data can contribute beyond genomics. Scenarios were evaluated using the metrics described below.}


To quantify the similarity between kernels built at different taxonomic levels, we use two complementary measures: (1) the RV coefficient, which is a multivariate generalisation of the squared Pearson correlation coefficient, 
providing a global measure of matrix similarity; (2) the $\kappa$ coefficient ~\cite{cuyabanoExpectedValuesAccuracy2024} which offers a graphical assessment based on the similarity of eigenvectors. 


We assess model quality along two dimensions: (1) phenotypic prediction accuracy, measured by the Pearson correlation between predicted $\hat{\mathbf{y}}$ and observed $\mathbf{y}$ phenotypes for the selection candidates, \textit{i.e.} individuals in the last generation;  and (2) variance component estimation to quantify the respective contributions of genetic and microbiota effects to the total phenotypic variance. One the one hand, phenotypic prediction accuracy is of practical relevance when the trait of interest drives selection \review{(specifically in plants)} or management decisions. 
One the other hand, accurate estimates of $b^2$ and $h^2_d$ are an important input for calibrating joint models such as MGBLUP, and more broadly for any BLUP-type model in which variance components are assumed to be known and directly condition the weights assigned to each source of information.

\section{Results}
\label{sec:RESULTS}

\begin{figure}[!h]
\vspace{3mm}
    \begin{center}
    \includegraphics[width=0.8\textwidth]{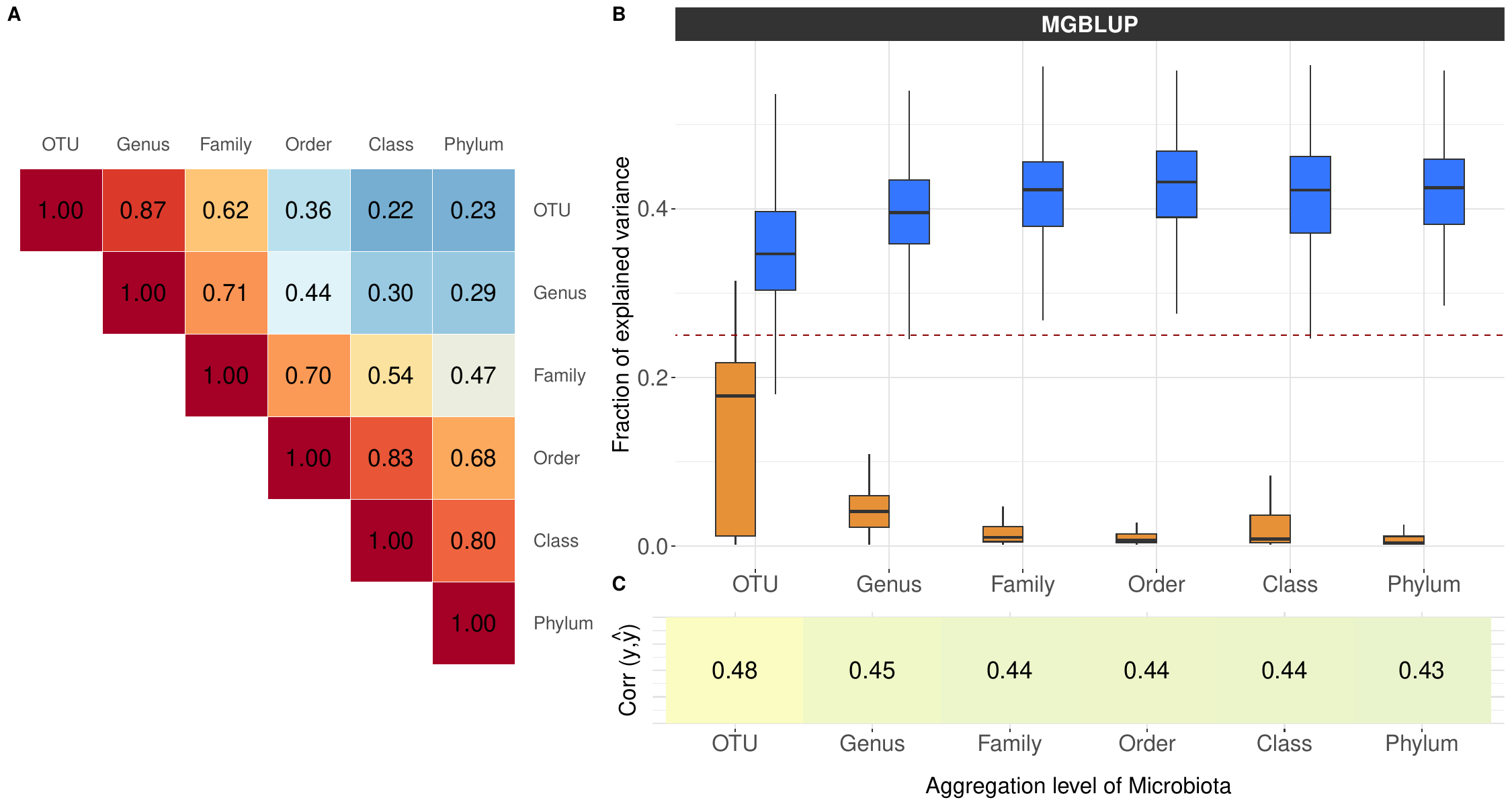}
    \caption{\label{figure:prediction_bis} \textbf{Comparison of similarity matrices and phenotypic prediction accuracy constructed using various levels of microbiota aggregation.} (A) the mean RV coefficent, averaged over 100 iterations, 
    (B) Microbiability ($b^2$, orange) and direct heritability ($h^2_d$, blue), estimated using simulated hologenomic data and an MGBLUP model, over 500 simulation replicates. The red dotted line corresponds to the true value ($b^2 = h^2_d = 0.25$). (C) Average correlation between real and predicted phenotypic values among selection candidates over 500 replicates.}
    \end{center}
\vspace{-8mm}
\end{figure}

\review{We compare similarity matrices computed at  various levels of taxon aggregation using the RV and $\kappa$ coefficients. We also investigate the impact of the granularity of taxon aggregation on variance component estimation ($b^2$ and $h^2_d$), and phenotypic prediction accuracy.} The mean RV coefficient between similarity matrices decreases rapidly with the taxonomic distance between the compared levels (Figure~\ref{figure:prediction_bis}.A). For example, the RV coefficient between OTU and genus levels is 0.87 and drops to 0.62 between OTU and family levels, indicating non-negligeable information loss induced by over-aggregation. Results based on the $\kappa$ coefficient, not shown here, led to similar conclusions.
Aggregating microbiota at increasingly higher taxonomic levels rapidly produces biased estimates of microbiability ($b^2$) and direct heritability ($h^2_d$) (Figure~\ref{figure:prediction_bis}.B). Phenotypic accuracy similarly declines when aggregating at increasingly higher levels beyond OTU and to a lesser extent genus-level (Figure~\ref{figure:prediction_bis}.C). However, predictive ability is largely maintained across all aggregation levels, reflecting the contribution of the genetic component that is correctly captured by the genomic kernel, regardless of microbiota resolution.

\begin{figure}[h!]
\vspace{3mm}
\begin{center}
    \includegraphics[width=0.9\textwidth]{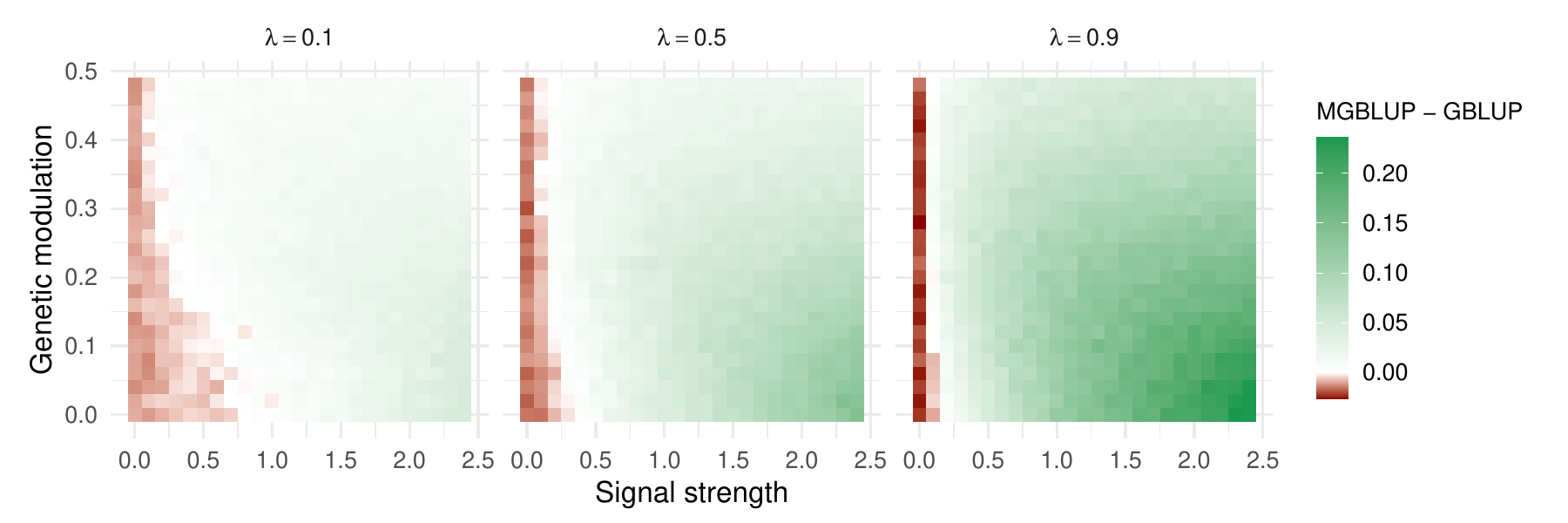}
    \caption{\label{figure:contour_plot_100} \textbf{Contour plot of the difference in phenotypic prediction accuracy between MGBLUP and GBLUP.} Each tile represents an average over 100 iterations. From left to right, panels correspond to an increasing proportion of the maternal contribution to the microbiota ($\lambda$ = 0.1, 0.5, 0.9). The x-axis corresponds to the relative importance of the microbiota on the phenotype compared to the direct genetic contribution ($b^2/h^2_d$), and the y-axis is the strength of the genetic modulation of the microbiota ($\sigma_\beta$). Hologenomic predictions can either underperform (red tiles) or outperform (green tiles) genomic prediction.}
\end{center}
\vspace{-8mm}
\end{figure}

Next, we investigate the added value of the microbiota, assessed as the difference in phenotypic prediction accuracy between GBLUP and MGBLUP, across contrasted scenarios and varied levels of maternal transmission $\lambda$, with microbiota abundances at the finest taxonomic resolution (OTU level), based on the above results. We focus on the joint impact of signal strength, quantified by the ratio $b^2/h^2_d$, and genetic modulation strength $\sigma_\beta$ on the added value of the microbiota (Figure~\ref{figure:contour_plot_100}). Three regions consistently emerge regarding this added value: (i) a detrimental region on the left, where weak microbiota contribution implies that adding the microbiota kernel is ineffective and introduces noise, (ii) a beneficial region in the lower right, where strong microbiota signal and moderate genetic control allow prediction gains, and (iii) a saturating region in the upper right, where no improvement is observed as the effect of microbiota is already captured by the genetic compartment due to strong modulation. Both the boundary between regions and the magnitude of gains increase with $\lambda$, as stronger maternal transmission structures the microbiota signal across generations.


\section{Discussion and conclusion}
\label{sec:CONCLUSIONS}

Our results demonstrate that the taxonomic level at which microbial abundances are aggregated critically influences variance component estimation. Two mechanisms account for this: first, pooling taxa with heterogeneous effects appears to dilute the true microbiota contribution to the phenotype. Second, aggregation alters the structure of the kernel matrix itself, leading to dissimilar matrices built at different taxonomic resolutions, as shown by the RV coefficients. These findings are in line with previous observations on the sensitivity of microbiome-based prediction to taxonomic aggregation ~\cite{armourGoldilocksPrincipleGut2022}.

Beyond aggregation, our simulations reveal that the added value of microbiota for phenotype prediction depends critically on the interplay between the relative contribution of microbiability $b^2$ compared to direct heritability $h^2_d$ and on the strength of host genetics on taxa abundances $\sigma_\beta$. When the microbiota is strongly influenced by the host genome, due to a high vertical transmission rate or strictly through host genetic modulation, its information is redundant with the genetic component, leading to only small gains or even slight losses in predictive ability when integrating it into the model. In contrast, in favourable scenarios, in which the microbiota both contributes a meaningful signal and is independent from the host genetics, integrating microbiota data improves phenotypic prediction by up to 20 points. Placing these results in a broader biological context, traits such as feed efficiency in livestock represent precisely the scenarios in which microbiota-informed models are expected to be most beneficial.

Two main limitations of the present work open avenues for future research. First, we relied exclusively on a linear kernel \review{and a single model class}, but exploring phylogenetic, sparse, or other biologically informed similarity matrices, \review{as well as alternative modelling approaches}, could better accommodate the structured nature of microbiota data. Second, \review{although our simulation was calibrated on real data, validation on empirical datasets remains necessary to confirm these findings in practice.} Third, we focused on phenotypic prediction; although this is informative, estimating breeding values is the central goal in animal breeding and a promising future study could capitalize on the true genetic values available within the RITHMS simulation framework to evaluate the added value of microbiota data for this purpose. \review{While this study illustrates these approaches in a livestock context, the underlying methodology is not restricted to animal breeding: the same framework could be applied to plant breeding or to human phenotypes, where the gut microbiome is similarly involved in complex traits.}

\section*{Conflict of interests}
\label{sec:CONFLICT-OF-INTERESTS}
\footnotesize \blue The authors have no competing interests to declare that are relevant to the content of this article. \black \normalsize

\section*{Acknowledgments}
\label{sec:ACKNOWLEDGMENTS}
\footnotesize \blue The authors acknowledge the HOLOFLUX metaprogram for labelling this work and for providing travel support. \black \normalsize 
\vspace{-8mm}
\section*{Funding}
\label{sec:FUNDING}
\footnotesize \blue This research is supported by the French National Research Agency under the France 2030 program ("ANR-22-PEAE-0006"), which supports future innovation in agricultural science.  \black \normalsize 

\section*{Availability of data and software code}
\label{sec:AVAILABILITY}
\footnotesize \blue Data and code used for this study are described and available in the following vignette : \scriptsize \url{https://solenepety.github.io/RITHMS/articles/hologenomic-data-for-phenotypic-prediction.html}). \normalsize \black

\footnotesize
\bibliographystyle{unsrt}
\bibliography{article2} 
\normalsize

\end{document}